\documentclass[11pt]{article}
\usepackage[normalem]{ulem}
\pdfoutput=1 % if your are submitting a pdflatex (i.e. if you have

\usepackage{jheppub} % for details on the use of the package, please
\usepackage{url}
\usepackage{caption}
\usepackage{subcaption}
\usepackage{extarrows}

\usepackage[latin9]{inputenc}
\usepackage{float}
\usepackage{amsmath}
\usepackage{amssymb}
\usepackage{graphicx}
\usepackage{esint}
\usepackage{hyperref}
\usepackage{comment}
\usepackage{color}
\usepackage{microtype}
\usepackage{cleveref}
\usepackage{breakurl}
\usepackage{bbm}
\newcommand{\be}{\begin{equation}}
\newcommand{\ee}{\end{equation}}
\newcommand{\ben}{\begin{displaymath}}
\newcommand{\een}{\end{displaymath}}
\newcommand{\bea}{\begin{eqnarray}}
\newcommand{\eea}{\end{eqnarray}}
\def\K{K{\"a}hler }
   \newcommand{\rf}[1]{(\ref{#1})}
\newcommand{\vp}{\varphi}

\def\be{\begin{equation}}
\def\ee{\end{equation}}
\def\bea{\begin{eqnarray}}
\def\eea{\end{eqnarray}}
\def\ba{\begin{array}}
\def\ea{\end{array}}
\def\bit{\begin{itemize}}
\def\eit{\end{itemize}}

\def\a{\alpha}

\def\vp{\varphi}
\def\m{\mu}

\newcommand{\N}{\mathcal{N}}
\newcommand{\cN}{\mathcal{N}}

 \makeatletter

\allowdisplaybreaks

\makeatother

\makeatletter
\DeclareRobustCommand{\rcite}[1]{%
  \rcite@aux#1,\@nil{#1}%
}
\def\rcite@aux#1,#2\@nil#3{%
  \if\relax#2\relax
    Ref.~\cite{#3}%
  \else
    Refs.~\cite{#3}%
  \fi
}
\makeatother

\hypersetup{
    colorlinks = true,
    citecolor = {blue},
    linkcolor = {blue},
    urlcolor = {blue},
}

 \title{\rm { \LARGE  \bf  Dilaton-Axion Inflation  with PBHs and GWs}}

\author{Renata Kallosh}
\author{ and Andrei Linde}
\affiliation{Stanford Institute for Theoretical Physics and Department of Physics,\\ Stanford University, Stanford, CA 94305, USA}
\emailAdd{kallosh@stanford.edu}
\emailAdd{alinde@stanford.edu}

\abstract{ We discuss  two-stage dilaton-axion inflation models  \cite{Linde:2018hmx}  and describe $\a$-attractor  models  with either  exponential or polynomial approach to the plateau.
We implement one of the  models of primordial black hole production proposed in \cite{Braglia:2020eai}  in the  $\a$-attractor context, and develop its supergravity version. The predictions of this model following from its polynomial attractor properties are:  $n_s$ and $r$ are $\a$-independent, $r$ depends on the mass parameter $\mu$ defining  the  approach to the plateau. The tachyonic instability at the transition point between the two stages of inflation is proportional to the negative curvature of the hyperbolic space $\mathcal{R}_K=-2/3\a$. Therefore
the masses of primordial black holes (PBHs) and the frequencies of small-scale gravitational waves (GWs)  in this model show significant dependence  on $\a$.  }

\begin{document}

\maketitle

% \tableofcontents{}
%\newpage
\parskip 5pt

\section{Introduction}

 Cosmological $\a$-attractors represent  a broad class of models  which can describe all presently available inflation-related observational data by a choice of a single parameter (or a single combination of two parameters) \cite{Kallosh:2013yoa,Ferrara:2013rsa,Kallosh:2022feu,Kallosh:2021mnu}.  These models can be formulated as models of a single real inflaton field $\vp$. However, a particularly good  theoretical motivation of these models is found in the framework of hyperbolic geometry based on $SL(2, \mathbb{R})$ symmetry  or $ SU(1,1)$ symmetry  of the kinetic terms. They arise naturally  in supergravity where the scalar field  is complex.  These models were called attractors because their cosmological predictions are rather stable with respect to considerable modifications of their potential. Many of these predictions are determined by the underlying hyperbolic geometry.

\begin{figure}[H]
\centering
\includegraphics[scale=0.21]{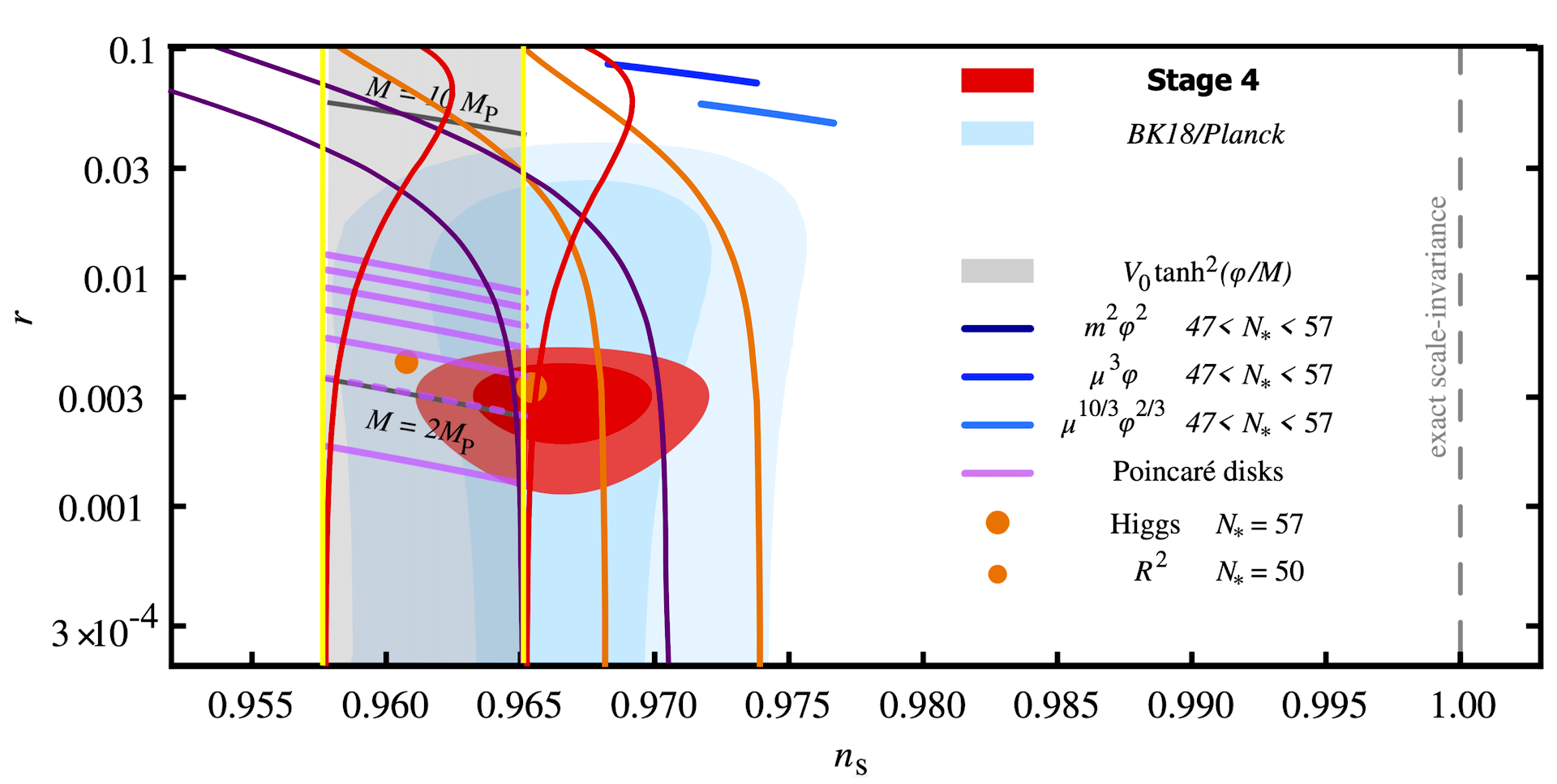}
\caption{\footnotesize  We use Figure 2 for  the $n_s-r$  plane  from `Snowmass2021 Cosmic Frontier: CMB Measurements White Paper' \cite{Chang:2022tzj}. Predictions of the simplest $\a$-attractor  T-models  with $V\sim \tanh^2(\vp/M)$ are shown in \cite{Chang:2022tzj} as a grey band. Seven light purple lines show predictions of advanced $\a$-attractor models inspired by string theory/M-theory models with  7 Poincar\'e disks.  We  added here also predictions of simplest E-models (area between two red lines), and predictions of quartic and quadratic polynomial KKLTI $\a$-attractors  (area between the dark purple and orange lines, respectively).}
 \label{SnowmassPlus}
\end{figure}

There are two simplest classes of such models: T-models, with potentials $V\sim \tanh^2(\vp/\sqrt{6\alpha})$, and E-models, with $V\sim (1-e^{-\sqrt{2/3\a}\,\vp})^{2}$, which predict $n_{s}= 1-2/N_{e}$ \cite{Kallosh:2013yoa}. In addition, there is a class of polynomial attractors \cite{Kallosh:2022feu}, which include KKLTI models with $V \sim {\phi^{2}\over \phi^{2}+m^{2}}$ with $n_{s}= 1-3/2N_{e}$ and $V \sim {\phi^{4}\over \phi^{4}+m^{4}}$ with $n_{s}= 1-5/3N_{e}$ \cite{Kallosh:2018zsi,Martin:2013tda}. As one can see in Fig. \ref{SnowmassPlus}, predictions  of  these simple single field inflation models completely cover the area favored by the latest Planck/BICEP/Keck data  \cite{BICEPKeck:2021gln}. These models can describe any small value of $r$, all the way down to $r = 0$.

Historically, in most of the models of this type, one of the two components of the complex field was stabilized during inflation, and the remaining one played the role of the inflaton.   In $SL(2, \mathbb{R})$  models the complex scalar is a dilaton-axion, the axion was usually stabilized, the dilaton was an inflaton. In $ SU(1,1)$ models the complex scalar represents  a Poincar\'e disk with its radial and angular components.  The angular component was stabilized, and the radial one played the role of the inflaton.

More recently  it was realized that models where both of the components  of the complex scalar field  contribute to the two-field dynamics of inflation may have some potentially interesting features.  
The supergravity versions of such models  were developed  in \cite{Kallosh:2015zsa,Achucarro:2017ing,Yamada:2018nsk} and  in \cite{Linde:2018hmx,Aragam:2021scu}. It was shown, in particular, that it is possible to find  supergravity description of models with arbitrary scalar potentials $V(T,\bar T)$  \cite{Achucarro:2017ing,Yamada:2018nsk,Linde:2018hmx}.  Bosonic versions of two-stage inflation in hyperbolic geometry, mostly in disk variables, were studied in  particular in \cite{Christodoulidis:2018qdw,Garcia-Saenz:2018ifx,Dalianis:2018frf,Anguelova:2020nzl,Iacconi:2021ltm,Pi:2021dft,Dalianis:2021dbs}; more references can be found in \cite{Aragam:2021scu}.

\begin{figure}[H]
\centering
\includegraphics[scale=0.2]{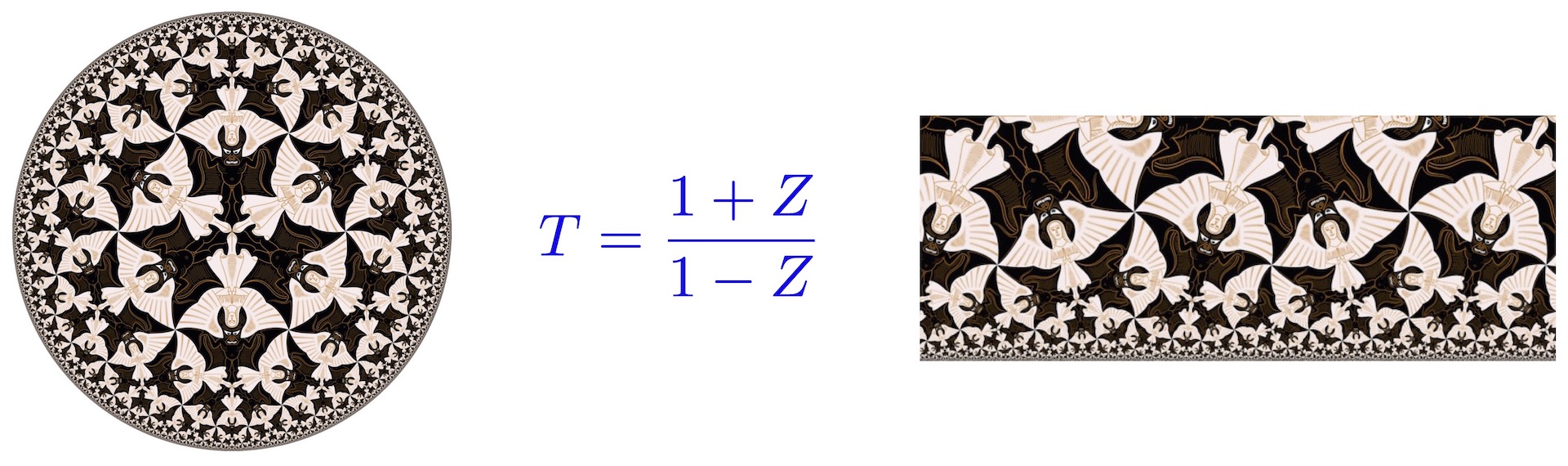}
\caption{\footnotesize Here are Escher's  pictures from \cite{Kallosh:2015zsa} for the hyperbolic space in disk coordinates, $|Z|^2<1$,  at the left and in half-plane coordinates, $T+\bar T>0$,  at the right. The boundary of the disk (with smaller and smaller angels and demons) is at $|Z|=1$, the boundary of the half-flat space is (with smaller and smaller angels and demons) at ${\rm Re} \, T=0$.
}
 \label{Escher2022}
\end{figure}

The relation between the dilaton-axion type half-flat space metric  and Poincar\'e disk metric describing hyperbolic geometry  was discussed in  the context of the $\a$-attractors in 
\cite{Kallosh:2015zsa,Garcia-Saenz:2018ifx,Iacconi:2021ltm}. All of these represent   negative space curvature manifolds with the \K curvature $\mathcal{R}_K=-2/3\a$. This was the way these models were introduced in \cite{Ferrara:2013rsa} in supergravity  with one chiral multiplet, i.e. one complex scalar\footnote{The curvature of the  hyperbolic geometry with 2 real fields is $\mathcal{R}=-4/3\a$.}. In  Fig. \ref{Escher2022} we show Escher's  pictures for the hyperbolic space in disk coordinates and in half-plane coordinates, from \cite{Kallosh:2015zsa}. In particular, in  the context of $\alpha$-attractors  \cite{Kallosh:2013yoa} the relation between the  T-models and E-models is the relation between Poincar\'e disk and half-flat space geometry: the kinetic terms are related by the change of coordinates, however,  the potentials are not.

Two-field inflationary models with non-trivial field metric have been used recently in the context of PBHs and GWs production. For single stage $\a$-attractors, the potential with inflection point\footnote{ Importance of the inflection point near the exit of inflation for PBH's production was realized in \cite{Garcia-Bellido:1996mdl,Garcia-Bellido:2017mdw,Ezquiaga:2017fvi,Germani:2017bcs,Ballesteros:2017fsr}.
} was introduced and studied in  \cite{Dalianis:2018frf,Iacconi:2021ltm,Dalianis:2021dbs} in disk coordinates. In   two-stage hyperbolic geometry  models, a far richer dynamical behavior is possible.

The new aspect of hyperbolic geometry, namely the metric in coordinates of the half-flat space taken in a standard stringy form of a dilaton-axion,  appears to play an important role when the primordial black holes,  gravitational waves  production, and reheating are studied. 
Various cosmological models with PBHs and GWs production and stages of reheating/preheating were investigated recently both in disk and well as in half-plane coordinates. Some of these, such as \cite{Iarygina:2018kee,Dalianis:2018frf,Anguelova:2020nzl,Iarygina:2020dwe,Iacconi:2021ltm,Dalianis:2021dbs}, were formulated in an obvious way as models in hyperbolic geometry, with curvature $\mathcal{R}=-4/3\a$, some others \cite{Braglia:2020eai,Pi:2021dft}  were not.

We will discuss here dilaton-axion inflationary models in hyperbolic geometry with exponential or polynomial $\a$-attractors and their supergravity realization. We will show that one of  the phenomenologically interesting dllaton-axion inflationary models proposed in \cite{Braglia:2020eai} can be interpreted as  a model with hyperbolic geometry. Then, using the methods developed in \cite{Kallosh:2015zsa,Achucarro:2017ing,Yamada:2018nsk,Linde:2018hmx}, we will develop the supergravity generalization of this model. 

This will allow us to use the $\alpha$-attractor predictions for $n_s$ and $r$ and compare them with the numerical results obtained in  \cite{Braglia:2020eai}. We will find a very good agreement. Moreover, our new understanding of inflation in \cite{Braglia:2020eai} will explain why  the PBH masses  and the frequencies of the GWs in this model depend on the curvature of the hyperbolic geometry. 

\section{2-moduli inflationary models in hyperbolic geometry}
\subsection{From exponential to polynomial $\a$-attractors}

 In the simple case the one-inflaton T- and E-models  \cite{Kallosh:2013yoa} with  {\it exponential approach to the plateau} have $n_s$ and $r$ independent of the properties of the large class of potentials.  However, $r$ depends on the  curvature of the \K geometry $\mathcal{R}_K= -{2\over 3 \a}$.  At large values of $\vp$ the potential is $ V_0( 1- e^{-\vp/\m} +\dots ) $. The tensor to scalar ratio $r$ depends on the parameter $\m = \sqrt {3 \a/ 2}$ describing the approach of the canonical field to the plateau.
\begin{figure}[H]
\centering
\includegraphics[scale=0.51]{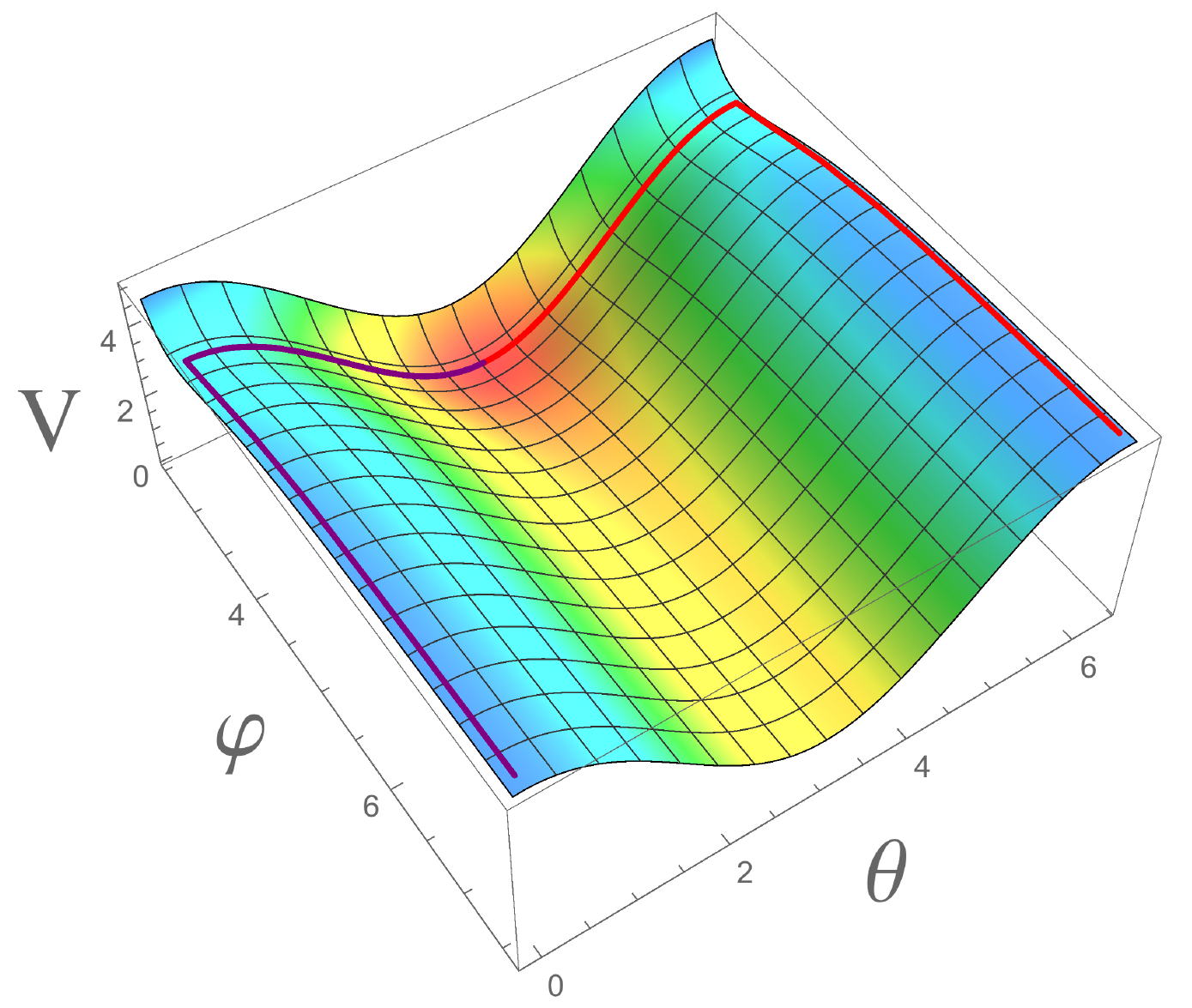}
\caption{\footnotesize One of the two-stage  dilaton-axion inflation models studied in \cite{Linde:2018hmx}. The dilaton plateau potential stage starts with some initial value of the axion and dilaton fields. Two different trajectories (purple and red) correspond to two different initial values of the axion  field. The axion during the dilaton inflation remains at its initial value, and starts moving only when the field $\phi$ approaches its minimum at $\phi = \phi_0$. Notice the sharp turn of the inflationary trajectories at $\phi = \phi_0$.}
 \label{PotT}
\end{figure}
In {\it polynomial  approach to the plateau}
one-inflaton $\a$-attractor models  \cite{Kallosh:2022feu}  the potential  at large values of $\vp$ is  $ V_0 \big ( 1- \big({\mu\over \vp}\big)^k +\dots \big )$. In these models $n_s$ depends on $k$, whereas $r$ depends both on $k$ and $\mu$, \cite{Kallosh:2018zsi,Martin:2013tda}. Both $n_s$ and $r$ do not depend on the  curvature of the \K geometry $\mathcal{R}_K=- {2\over 3 \a}$. 

Here we  will describe the  two-stage  dilaton-axion models of inflation developed in \cite{Linde:2018hmx}, as well as their their generalizations.   The original E-model dilaton potential with the exponential approach to plateau studied in \cite{Linde:2018hmx} can be  replaced by a polynomial attractor \cite{Kallosh:2022feu}, i.e. we can replace an E-model  potential by the one which has a power law approach to the plateau, a quadratic KKTTI attractor \cite{Kallosh:2018zsi,Martin:2013tda}.

Both types of models have the same dilaton-axion kinetic term 
\be
{\cal L}_{\rm kin} ={1\over 2} \Big [ (\partial \vp)^2  + e^{2 \sqrt {2\over 3\alpha} \vp} (\partial a)^2 \Big ] \ .
\label{kin}\ee
The  potentials studied in \cite{Linde:2018hmx} are
\be
V(\vp, a)= V_0 \Big( 1- e^{-\sqrt{2\over 3\a} (\vp - \vp_0)} \Big ) ^2 + {1\over 2} m_a^2 a^2 \ .
\label{HN}\ee
The dilaton potential in \rf{HN} is the E-model  exponential $\a$-attractor  with the position of the minimum  at $\vp=\vp_0$. The axion has  a mass term potential\footnote{The  axion potential used in  \cite{Linde:2018hmx}  at $\phi = \phi_0$ is more general, it is  proportional to $\cos^2  {\theta\over 2}$ where $\theta -\pi=\sqrt{2\over 3\a} a$. For small deviation of the axion from the minimum the model becomes the one shown in \rf{HN}.}.

We now replace the potential $V_{0}\Big( 1- e^{-\sqrt{2\over 3\a} (\vp - \vp_0)} \Big ) ^2$ with an exponential approach to the plateau used in \cite{Linde:2018hmx}  by a  polynomial attractor \cite{Kallosh:2022feu}, 
 which has a power law approach to the plateau, and find the dilaton-axion potential of the form
\be
V(\vp, a)=V_0 {\vp^2\over \vp^2+ \m^2} +{1\over 2} m_a^2 a^2 \ .
\label{Br}\ee
%***We could also study a more general potential with a minimum at any value of $\vp_{0}$, but in the context of this simple model this change can be absorbed into a redefinition of the axion mass.***

In a class of models considered in \cite{Braglia:2020eai,Pi:2021dft}, where the PBHs can be produced,  a kinetic term was chosen in the form
\be
{\cal L}_{\rm kin} =  {1\over 2} \big[ \partial_\mu \vp \partial^\mu \vp  +f (\vp) \partial_\mu a \partial^\mu a \big] \ .
\ee
One of the models studied in \cite{Braglia:2020eai} has 
\be
f(\phi) = e^{2 b_1 \vp} \ ,
\label{Bra}\ee
 but it was not clear whether this choice may have a  fundamental geometric interpretation, and whether it is any better than another choice $f(\phi) = e^{2 b_2 \vp^{2}}$ also studied in  \cite{Braglia:2020eai}.

Now we can relate it to the hyperbolic geometry of $\alpha$-attractors and incorporate this model in  supergravity with the corresponding \K geometry  iff
\be
f(\phi) = e^{2 b_1 \vp}= e^{ 2 \sqrt {2\over 3\alpha} \vp}  \, ,  \qquad  {\rm i.e.}   \qquad b_1 = \sqrt {2\over 3\alpha} \ .
\label{curv}\ee

Thus, if we  choose the kinetic term  in \cite{Braglia:2020eai} as in \rf{Bra},
we can  relate their parameter $b_1$  to the \K curvature, 
\be
b_1^2 = - \mathcal{R}_K= {2\over 3\alpha}\, ,  \qquad e^{2 b_1 \vp}= e^{2 \sqrt {|\mathcal{R}_K|} \vp} \ .
\ee

The phenomenology of this cosmological model introduced  in  \cite{Braglia:2020eai}, which  we now embedded into hyperbolic geometry,   was studied extensively with regard to PBHs and GWs. We will describe the microscopic origin of both models in \rf{HN}, and in \rf{Br} in supergravity,  and extract an interesting information about phenomenology from the fact that the model in \rf{Br} is a dilaton-axion inflationary model in hyperbolic geometry which has attractor properties of polynomial $\a$-attractors.

Importantly, one of the parameters in this model defining the properties of PBHs and GWs is actually a curvature of the hyperbolic geometry, as shown in   \rf{curv}. Other features of this model following from the fact that it is embedded into hyperbolic geometry with inflationary plateau potential approaching polynomially  \cite{Kallosh:2022feu} will be identified. We will show that  $n_s$ and $r$ are defined by their plateau potential attractor values \cite{Kallosh:2018zsi,Kallosh:2022feu}. 

The point of transition to the second stage of inflation is characterized by the effective mass square of the isocurvature perturbations $m^2_{\rm eff}$. It can abruptly and   temporarily become  large and negative; we will find out that  $m^2_{\rm eff} \sim \mathcal{R}_K$.

\subsection{Hyperbolic geometry in half-plane coordinates}

A simple way to introduce this class of metrics is to start with equation (5) in \cite{Kallosh:2007ig}, where the \K potential defining the \K geometry metric  is given by 
\be
K= - 3\alpha \ln (T+\bar T)\, ,
\ee
in notation  ($c=3\alpha$) adapted to the   definition of $\a$-attractor E-models. The \K metric is
$g_{T\bar T} = \partial _T \partial_{\bar T}  K$ and the geometry defines the kinetic term for a complex scalar 
\be
{\cal L}_{\rm kin} = 3\alpha {\partial_\mu T \partial_\nu  \bar T g^{\mu\nu}\over (T+\bar T)^2} \ .
\label{Emodel}\ee
It has an $SL(2, \mathbb{R})$ symmetry. It was explained around equations (5) and (6) in \cite{Kallosh:2007ig}   that the case $3\alpha=3$ corresponds to the  dilaton-axion in string theory \cite{Kachru:2003aw} where the total volume is defined by the dilaton, and no-scale supergravity \cite{Cremmer:1983bf,Ellis:1983sf}. The case  $3\alpha=1$ is a single dilaton-axion case. Now we know more examples of discrete $3\alpha =7,6,5,4,3,2,1$ \cite{Ferrara:2016fwe} as well as continuous $3\a$ in the context of $\cN=1$ supergravity \cite{Ferrara:2013rsa,Kallosh:2013yoa}. Following  \cite{Kallosh:2007ig} we present our complex field $T$, a half-plane coordinate of the hyperbolic geometry,  as a dilaton and an axion:
\be
T(x) = e^{- \sqrt{2\over 3\alpha} \vp (x) } + i \sqrt {2\over 3\alpha} \, a(x) \,  , \qquad T+\bar T>0 \ .
\label{T}\ee
The real part of the $T$-field is an exponent of the dilaton $\vp$, therefore clearly positive, which explains the `half-plane' coordinate name.
The dilaton-axion kinetic term \rf{Emodel} is
\be
{\cal L}_{\rm kin} = {1\over 2} \Big [ \partial_\mu \vp \partial_\nu \vp g^{\mu\nu} + e^{2 \sqrt {2\over 3\alpha} \vp} \partial_\mu a \, \partial_\nu a g^{\mu\nu}\Big ].
\label{da}\ee
Here $\vp$ is a dilaton and $a$ is the axion whose kinetic term couples exponentially to the dilaton.
 \be
  \vp= -  \sqrt{3\alpha \over  2} \ln {1\over 2 } (\bar T +T )   \, ,  \qquad  a= {1\over 2 i} \sqrt{3\a\over 2} (\bar T -T )   \ .
   \ee
   The total non-gravitational Lagrangian, including the potential, in geometric variables  is
   \be
{\cal L}  = 3\alpha {\partial_\mu T \partial_\nu  \bar T g^{\mu\nu}\over (T+\bar T)^2}  -  V(T, \bar T) \ .
\label{total}\ee

   \subsection{Supergravity version of the 2-moduli models }
The supergravity version of the hyperbolic geometry models with both dilaton and axion evolving during inflation was developed  in \cite{Kallosh:2015zsa,Achucarro:2017ing,Yamada:2018nsk} and \cite{Linde:2018hmx}. In addition to the dilaton-axion multiplet $T$  we need a nilpotent one, $X^2=0$. The \K potential and the superpotential for models of our interest here were proposed and studied in \cite{Yamada:2018nsk} and in \cite{Linde:2018hmx}. 

We generalize  supergravity models in \cite{Yamada:2018nsk} and in \cite{Linde:2018hmx} by introducing two different parameters. One is describing the breaking of supersymmetry in the $X$ directions,  we call it $F_X$, the other describing the breaking of supersymmetry in the $T$ directions will be $W_0$ as in  \cite{Yamada:2018nsk} and in \cite{Linde:2018hmx}. Our model is now defined as follows, for $\alpha<1$
\be
K= - 3\alpha \ln (T+\bar T) +G_{X\bar X} X\bar X\, ,  \qquad  W= W_0 + F_X X \ .
\label{KW}\ee 
\be
G_{X\bar X} = {F_X^2 \over  (T +\bar T) ^{3\alpha} V(T, \bar T) + 3W_0^2 (1-\alpha)} \ .
\label{G}\ee
The value of the potential in this model, at $X=0$ is
\be
V_{\rm final} = V(T, \bar T) \ .
\label{final}\ee

Thus, in terms of the dilaton and axion bosonic fields of our supergravity models \rf{KW}, \rf{G}, the non-gravitational part of the action is
\be\label{full}
{\cal L}(\vp, a)  = {1\over 2} \large[ (\partial \vp)^2 + e^{2 \sqrt {2\over 3\alpha} \vp} (\partial a )^2  \large] - V\big (T (\vp, a) , \bar T (\vp, a)\big )\, , \quad (\partial \vp)^2\equiv  \partial_\mu \vp \partial_\nu \vp g^{\mu\nu} \ .
\ee
This expression is the same as the one in eq. \rf{total}, for any potential $ V(T, \bar T)$. 
\subsection{Dilaton-axion models  in manifestly geometric variables}

Now we can present  these models in geometric hyperbolic variables $T$ where the action is given in eq. \rf{total} 
and  $T$ is defined in eq. \rf{T}. Using
\be
{1\over 2 } (\bar T +T ) \equiv t \ ,  \qquad  {1\over 2 i}  \sqrt{3\a\over 2}(\bar T -T ) \equiv a  \ ,
\label{ta}\ee
we can present the potentials of the models in \cite{Linde:2018hmx,Braglia:2020eai} which can be used for the supergravity version of the model.

The supergravity version of the Hypernatural inflation model \rf{HN} requires a potential depending on geometric fields $T, \bar T$. It is given by
\be
V(T, \bar T)= V_0  \big ( 1-t\big )^2 +  {1\over 2} m_a ^2  a^2 \ .
\ee
where $t(T, \bar T) $ and $a(T, \bar T)$ are given in eq. \rf{ta}. Replacing $t$ by the canonically normalized field $\vp$ results in \rf{HN} \cite{Linde:2018hmx}.

The cosmological model developed in \cite{Braglia:2020eai}   is
 \be
{\cal L}= {1\over 2} \bigl[ -(\partial_t \phi)^2  - e^{2 b_1 \phi} (\partial_t \chi)^2 \bigr  ] - V_0 {\phi^2\over \m^2 + \phi^2} - {1\over 2} m_\chi ^2\chi^2  \ .
\label{finelli}\ee
 It is now easy to recognize it as the one we presented in eqs. \rf{kin}, \rf{Br}
 under condition
\be
b_1= \sqrt {2\over 3\alpha}\, , \qquad 
 \vp \rightarrow  \phi \, , \qquad 
a  \rightarrow  \chi \ .
 \ee
It can also be given in the form with the  $SL(2, \mathbb{R})$ invariant metric  \rf{Emodel}.
The potential breaks the $SL(2, \mathbb{R})$ symmetry of the kinetic term and is given by the expression where the  inflaton part of the potential was presented in \cite{Kallosh:2022feu}. Namely, 
the potential of model \cite{Braglia:2020eai}  in geometric  variables is
\be
V(T, \bar T) = V_0 {\ln^2 t \over c^2+ \ln^2 t} +  {1\over 2} m_a ^2  a^2 \, , \qquad c^2= \mu^2 |\mathcal{R}_K|  \ .
\label{Pott}
\ee
where $t(T, \bar T) $ and $a(T, \bar T)$ are given in eq. \rf{ta}.
This also means that the supergravity version of the model in \cite{Braglia:2020eai} with \rf{Bra}  is now available in eqs. \rf{KW}, \rf{G}, \rf{final},  \rf{Pott}.  It is a dilaton-axion cosmological attractor model of the type described in \cite{Kallosh:2022feu}.
\begin{figure}[H]
\centering
\includegraphics[scale=0.60]{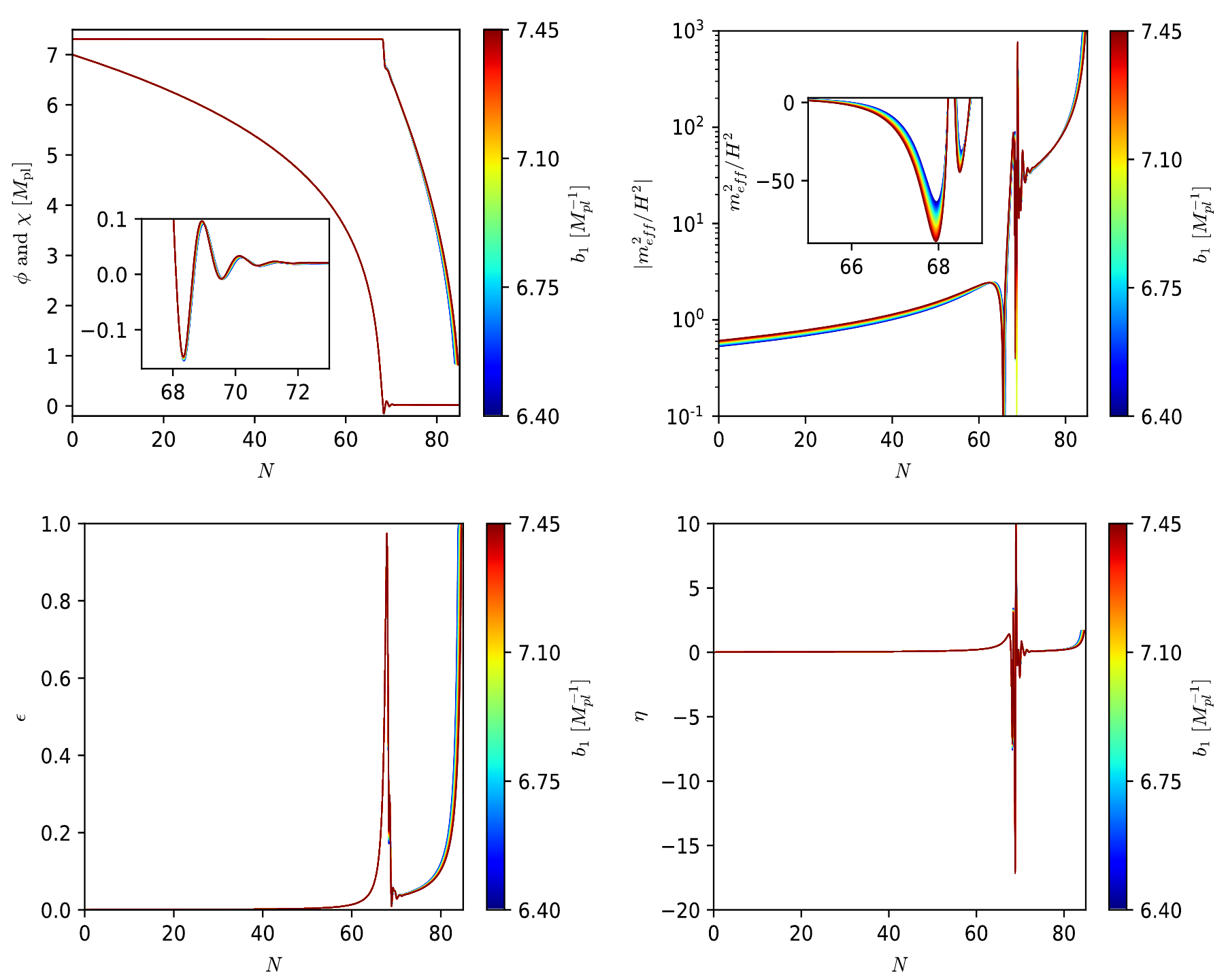}
\caption{\footnotesize The cosmological predictions of a model in \cite{Braglia:2020eai},  Fig. 1 there based on numerical solution. The figures show the  evolution of the dilaton and axion, of the slow-roll parameters and the effective mass of the isocurvature perturbation. The dependence on $\a$, which here is the dependence on the color codifying $b_1= \sqrt {2\over 3\alpha}$, is practically absent. This is in agreement with attractor eqs. \rf{nsr1}.}
 \label{Finelli3}
\end{figure}

\section{Observational predictions of the axion-dilaton attractor}

\subsection{CMB predictions for $n_s$ and $r$}

The $\a$-attractor models with stabilized axion have stable attractor predictions. In case of the polynomial $\a$-attractors which we have here, we can first look at numerical examples studied in \cite{Braglia:2020eai} and see if they are supported by the  attractor values presented in eq. (2.14) in \cite{Kallosh:2022feu}. 
For slow roll parameters we have
\be 
n_{s} =  1-{3\over 2N_{e}} ,\qquad r  = {\sqrt 2 \m \over N_{e}^{3/2}}  \ . 
\label{nsr1}\ee
We compare this  prediction with the numerical cases studied in  \cite{Braglia:2020eai} and displayed in their Fig.~1, we show them in our Fig. \ref{Finelli3}.
\begin{figure}[H]
\centering
\includegraphics[scale=0.37
]{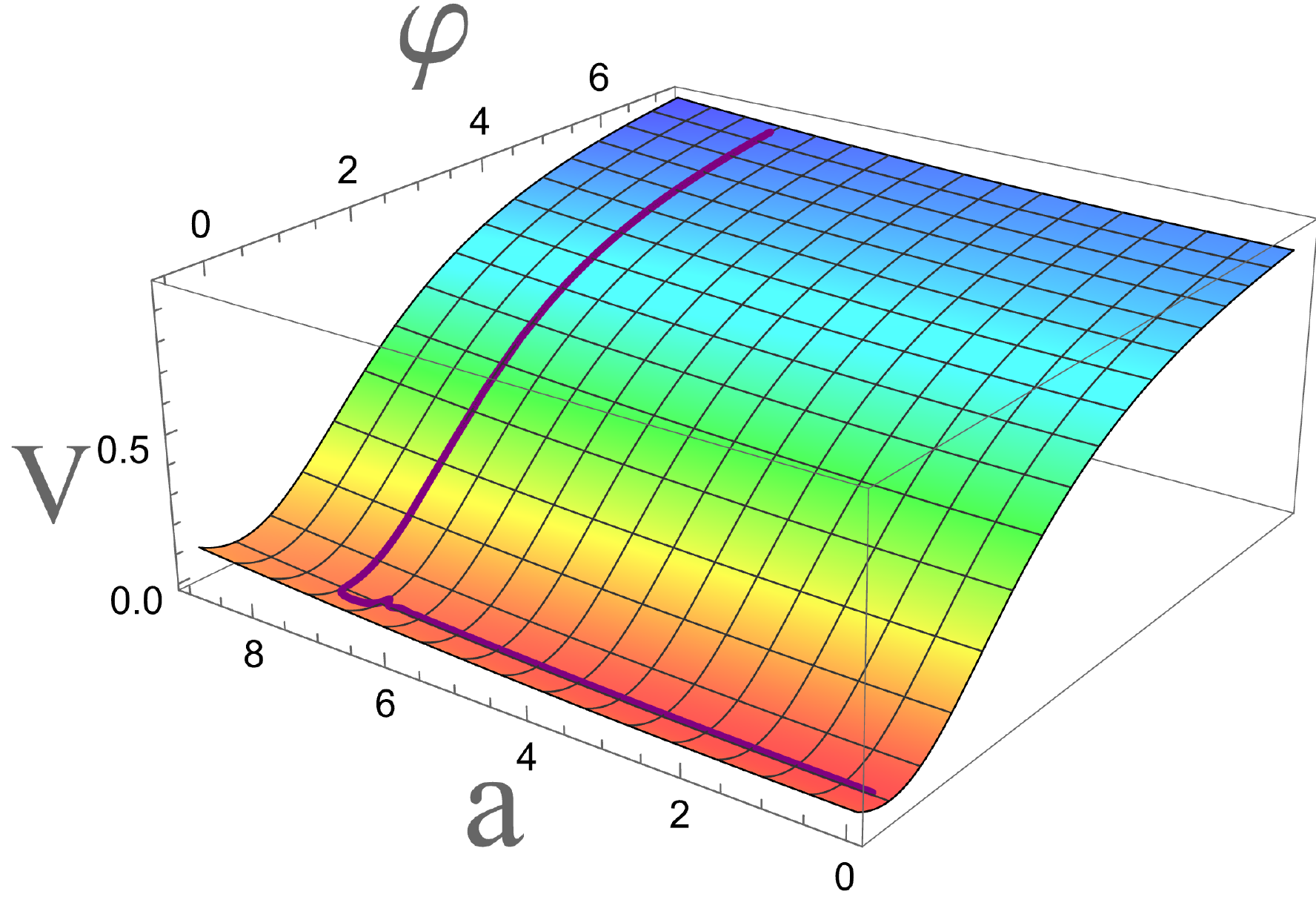}
\caption{\footnotesize The dilaton-axion potential in terms of a dilaton which has a canonical kinetic term and the axion $a$ which has an exponential coupling to dilaton in his kinetic term.  Parameters are from  \cite{Braglia:2020eai}: $\mu =\sqrt 6$, ${V_0\over  500}=m_a^2 $, and we took $\a=0.0136$ for $b_1=7$. The  dilaton stage of inflation happens when the dilaton field rolls down the blue plateau of the KKLTI potential, The second stage is the axion chaotic inflation due to the  quadratic potential shown as the  red area. }
 \label{PotT2}
\end{figure}
We can explain the main features of the model based on its attractor properties as well as numerical solutions  supporting them, see Figs. \ref{Finelli3}, \ref{PotT2}. The effective mass of the axion is suppressed by the exponential  factor $e^{- \sqrt {2\over 3\alpha} \vp}$ due to kinetic term coupling.  The axion is exponentially light at large $\vp$ at the plateau, and starts moving only when the field $\vp$ approaches the minimum of its potential at $\vp = 0$, and  the exponential suppression of its effective mass disappears. This is the `rolling on the ridge effect' effect  found  in \cite{Linde:2018hmx,Achucarro:2017ing}.  Thus the dilaton stage of inflation ends at $a\approx a_{i}$.
After $\vp$ reaches $\vp=0$, the axion in models with $a_{i} \neq 0$,  undergoes a stage of chaotic inflation due to the axion potential
$
{1\over 2} m_a ^2  a^2
$. The number of e-folds at this stage depends on how far is $a_{i} $ from its minimum $a=0$ when it rolls along the  valley with the quadratic potential shown as the elongated  red area at $\vp = 0$ in Fig. \ref{PotT2}. We plot the trajectory of the fields during the two-stage inflation.
At the turning point we  found that $\epsilon$ jumps to $1$ and $\eta$ jumps down to $-17$ and rises up to $+10$ before settling to the slow roll axion inflation stage.

For the total number of e-folds to be 57, the number of e-foldings $N_e$  responsible for observations in eq. \rf{nsr1} should be  smaller than 57, in view of additional e-folds at the axion stage:
\be
N_{\rm total} = N_e + N_{\rm axion} \approx 57 \ .
\ee
Note that if initially at $\phi= 7$ the axion would be placed at $a=0$, in the middle of the plateau, it would be only one stage of inflation with $N_e=57$, $n_s \approx 0.97 $. 

\begin{figure}[H]
\centering
\includegraphics[scale=0.75]{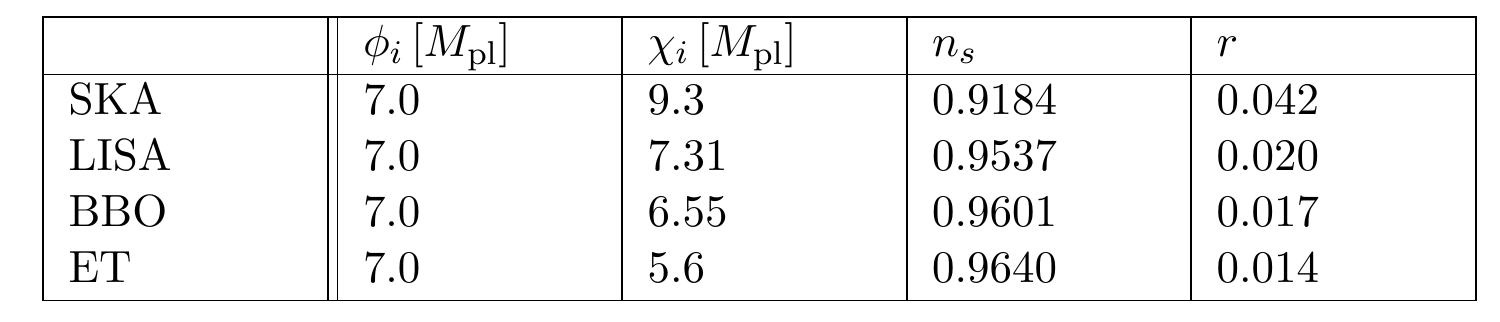}
\caption{\footnotesize Initial conditions for the dilaton and axion and $n_s$ and $r$ .}
 \label{Table}
\end{figure}

We will consider 4 different cases which differ by the choice of the initial position of the axion, as we see in   Table 1 in \cite{Braglia:2020eai}  and reproduced   here in   Fig. \ref{Table}.  
The number of e-foldings $N_e$ which we have to use in our attractor formula \rf{nsr1} can be identified by the values of $n_s$ or $r$ in the table.  We get $N_e$ from $r$ using the table, and calculate $n_s$ with the same $N_e$ we find a  good agreement with the table of numerical solutions given in  \cite{Braglia:2020eai}. Namely, we find
for SKA, LISA, BBO, ET examples from \rf{nsr1}
\bea
N_e=19   &&  n_s= 0.921 \qquad r=0.042\cr
N_e=32  && n_s= 0.952 \qquad r=0.020\cr
N_e=35  && n_s= 0.957 \qquad r=0.017\cr
N_e=39  && n_s= 0.961 \qquad r=0.014
\eea
One should take into account also that the  parameters  $\m=\sqrt 6$ and $r > 10^{-2}$  are above the values of these parameters where the attractor regime is reached in these models, as explained in \cite{Kallosh:2018zsi}.  One can also see it in Fig. 3 in \cite{Kallosh:2022feu}, which shows that the attractor regime is reached only for $r  \lesssim 10^{-2}$. 

And we definitely see the trend which is also clear from the properties of the dilaton-axion model of inflation. In particular, the attractor values of $n_s$ and $r$ in eq.  \rf{nsr1} are $\a$-independent, which
clearly  explains  why in Fig. \ref{Finelli3} almost all curves for different values of $b_1=\sqrt {2\over 3\alpha}$ coincide.

\subsection{PBHs and GWs dependence on hyperbolic space curvature}
\begin{figure}[H]
\centering
\includegraphics[scale=0.52]{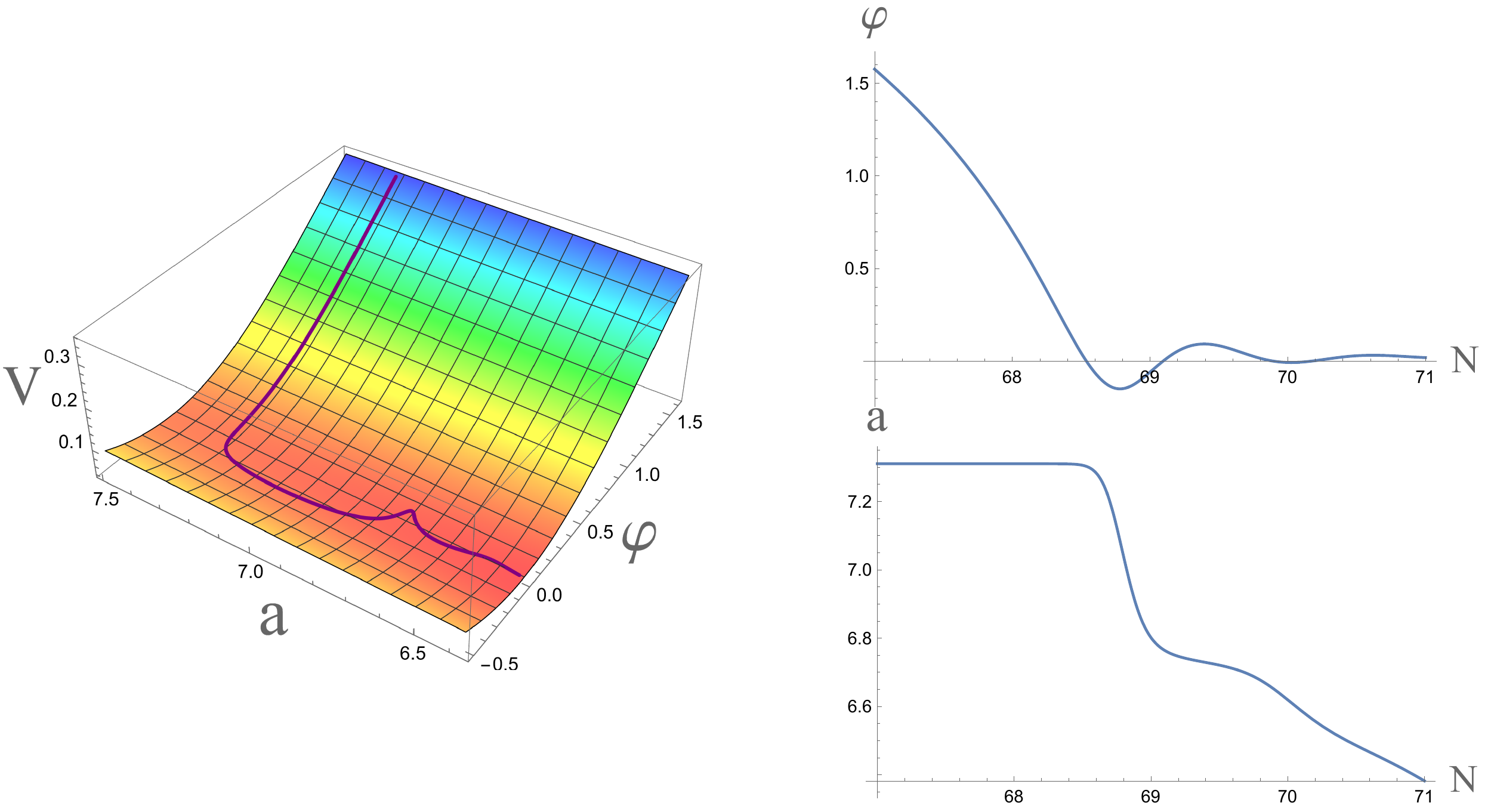}
\caption{\footnotesize Near transition area where the axion inflation stage replaces the dilaton inflation.  We can see the correlation between the fast change of the axion and $\vp$ crossing zero and oscillating to negative value.}
 \label{m23}
\end{figure}

Here in Fig. \ref{m23} we describe the transition area at the fixed value of \K curvature. The kinetic term for the axion field is $e^{2\sqrt {| \mathcal{R}_K | }\vp}(\partial a)^2$. It means that the `physical distance' is  $e^{\sqrt {| \mathcal{R}_K | }\vp}\partial a$, which is the reason why at large positive $\vp$ the axion field $a$ is not moving for a long time. We can see it in Fig. \ref{Finelli3} at the upper left corner where $\chi=a$ remains at his initial position for a long time during the first stage of the dilaton inflation. When the field $\vp$ reached his minimum at $\vp=0$ this protection of the axion position at its initial value vanishes. Moreover, when $\vp$ becomes negative during the oscillation near the minimum, this factor becomes $e^{ -\sqrt {| \mathcal{R}_K | } |\vp |} \partial a$. This forces the axion first to change dramatically from its initial position and after a while a slow roll axion inflation takes place.
  Note that this effect is significant for large curvatures $| \mathcal{R}_K | $ and less significant for smaller ones, which explains the color dependence of the value of $m^2_{\rm eff}/H^2$ in the blow-up region in the upper right plot in Fig. \ref{Finelli3}. The color there is codifying $b_1= \sqrt {2\over 3\alpha}$, it changes from blue to red when $b_1$ is increasing. Note that in all figures in in Fig. \ref{Finelli3} there is no color dependence. 
  The only exception  is a blow-up region in the upper right plot of the effective mass of isocurvature perturbations $m^2_{\rm eff}/H^2$. This color-dependence of $m^2_{\rm eff}/H^2$ is practically absent during the first and the second stage of inflation, it is only present at the  point of transition. 
Our Fig. \ref{m23}  compliments a blow-up region in the upper right plot of the effective mass of isocurvature perturbations $m^2_{\rm eff}/H^2$ in Fig. \ref{Finelli3}. 
The effective mass of isocurvature perturbations $m^2_{\rm eff}$ becomes temporarily negative at the transition between the two stages of inflation and leads to a transient tachyonic amplification of the isocurvature perturbations  leading to a large peak in the power spectrum. 
 We found an additional explanation of the color dependence of $m^2_{\rm eff}$ defined in  \cite{Braglia:2020eai} at the region of transition where $\phi_{tran}\approx 0$. We have found that the effective mass to Hubble ratio depends on the \K curvature $ \mathcal{R}_K = - b_1^2$  as follows
 \be
 { m^2_{\rm eff}\over H^2} \Rightarrow  5\,   \mathcal{R}_K   \Big ({\partial a \over \partial N} \Big )^2 + \dots 
\label{mass} \ee
The  $\dots$  in \rf{mass}  are  for terms which mildly depend on $\mathcal{R}_K$, only in the form
$e^{\sqrt {| \mathcal{R}_K | }\phi_{tran} }$ with $\phi_{tran}\approx 0$,  or are independent on $\mathcal{R}_K$.

 The first term is linear in \K curvature, depends on how fast the axion $a$ changes as a function of $N$,  and it is dominant at the  transition, the second term depends on \K curvature times $\phi_{tran}\approx 0$. This means that at the region  of transition the negativity of the effective isocurvature mass is due to the negativity of the \K curvature  as the first term in eq. \rf{mass} shows. 
 
 \begin{figure}[H]
\centering
\includegraphics[scale=0.50]{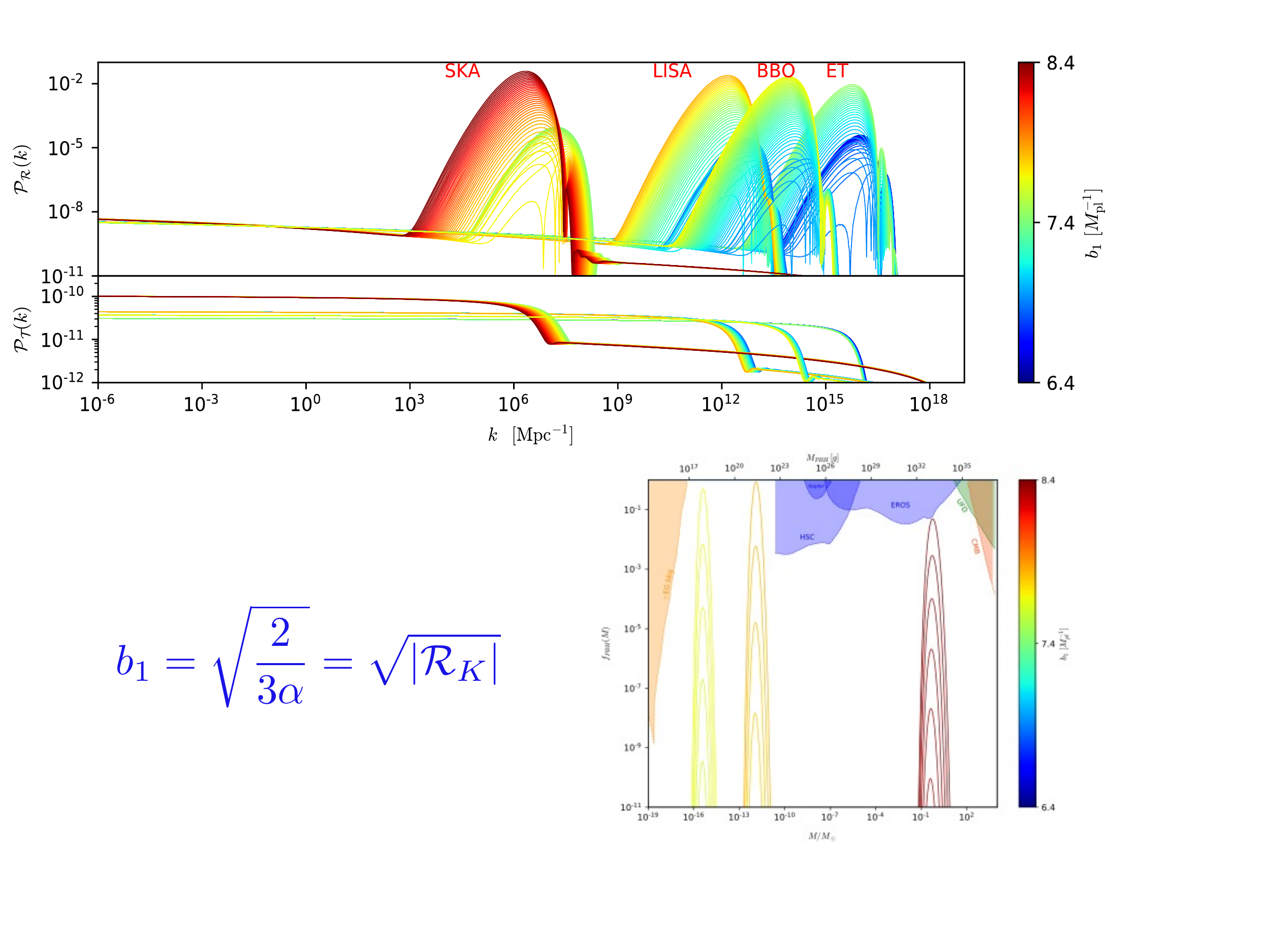}
\caption{\footnotesize The cosmological predictions of a model  in \cite{Braglia:2020eai} where the  potential with polynomial approach to plateau is embedded into hyperbolic geometry has $\a$-dependent  predictions  for the  the tensor power spectra at the end of inflation and for the properties of the PBHs.
}
 \label{Finelli}
\end{figure}
\begin{figure}[H]
\centering
\includegraphics[scale=0.50]{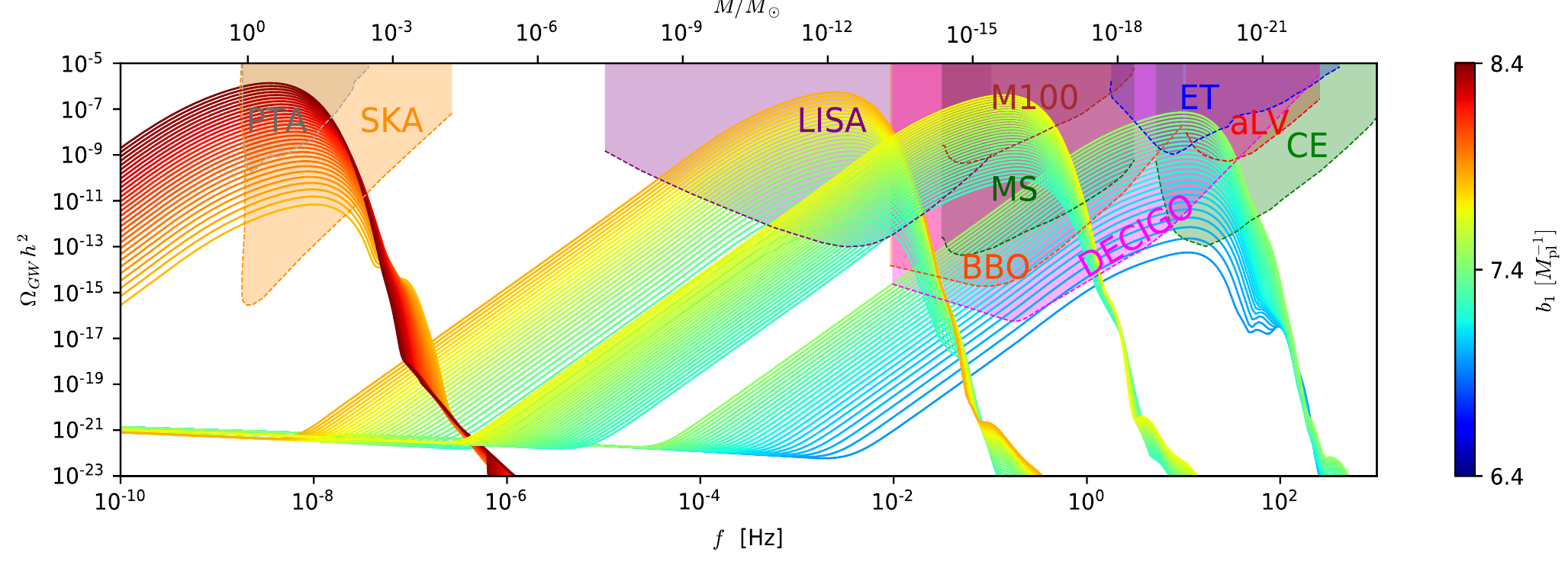}
\caption{\footnotesize The cosmological predictions of a model in \cite{Braglia:2020eai} where  $b_1$-dependence of   the properties of the induced GW is presented by color coding and $b_1= \sqrt {2\over 3\alpha} $.
}
 \label{Finelli2}
\end{figure}

The upper right plot
in Fig. \ref{Finelli3} shows that the most negative effective mass of isocurvature perturbations in the  blow-up region  can be explained by the formula \rf{mass} which says that for bigger $|\mathcal{R}_K | $ the most negative value is reached (the red part of the figure at  $N\approx 68$). This property of the   ratio $m^2_{\rm eff}/H^2$ explains why
 the phenomenology of the PBHs and GWs strongly depends on the curvature.
In Figs. \ref{Finelli} and  \ref{Finelli2} we reproduce figures from \cite{Braglia:2020eai} where one can see that the change in $b_1=\sqrt {2\over 3\alpha}$ leads to dramatic changes in phenomenology. 
In particular, when $ 6.4\leq  b_1 \leq 8.4$ we need very small $\a$ so that $0.028 \leq 3\alpha \leq 0.049$ and very high \K curvature $-\mathcal{R}_K =  b_1^{2} \gtrsim 40$.

\begin{figure}[H]
\centering
\includegraphics[scale=0.65]{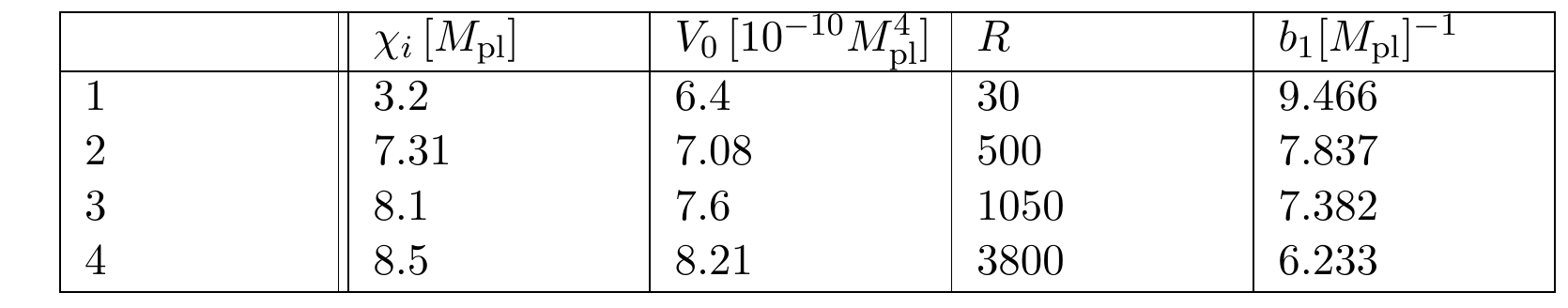}
\caption{\footnotesize Changes in axion masses $m_a^2=  {V_0 \over R} $ and initial conditions for the axion and the resulting change in \K curvature $b_1^2={2\over 3\alpha} = - \mathcal{R}_K $.}
 \label{Table2}
\end{figure}

An additional set of models with parameters ${V_0\over m_a^2}= R=30, 1050, 3800$ was also studied in  \cite{Braglia:2020eai}, see Figure \ref{Table2} here.

It was observed in \cite{Aragam:2021scu} that rapid-turn inflation models in hyperbolic geometry tend to have  high-curvature. This underscores the difficulty of obtaining such models from string theory. For example, the ones associated with M-theory and type IIB string theory  in \cite{Kallosh:2021vcf} without rapid terms have $3\a= 7,6,5,4,3,2,1$.

Interestingly, with increasing $R={V_0\over m_a^2}$ which corresponds to lighter axions, the required value of  $b_1$ is decreased and $\a$ is increased.  We can provide here a qualitative explanation of this feature, comparing the cases with $R=500$ and $R=3800$ in \cite{Braglia:2020eai}.

The mass squared of the axion is decreasing  from the case described in details with $m_a^2=  {V_0 \over R} $ to the case $ \tilde m_a^2$ with different $V_0$ and $R$ and (case 2 and case 4 in Fig. \ref{Table2}).  Also the initial conditions change from $a_i$ to $\tilde a_i$. We can compare the effective mass of the axion in case 2 and case 4
\be
e^{-2b_1 \phi} m_a^2 a_i^2\, ,  \qquad e^{-2\tilde b_1 \tilde \phi} \tilde m_a^2 \tilde a_i^2 \ .
\ee
It follows that 
\be
e^{2 b_1 \phi_c - \ln C^2} \approx e^{2 \tilde b_1 \tilde \phi_c} \ .
\label{newa}\ee
and numerically $\ln C^2\approx 1.57$.
If we assume that we can compare these two models at the point $\phi_c\approx 0.49$  which is approximately the same in both cases, we find that 
\be
b_1-\tilde b_1\approx 1.55 \ ,
\ee
to be compared with the difference between $b_1= 7.837$ to $\tilde b_1= 6.233$, which is $1.6$.

This is desirable modification of the parameters in the model since increasing $\a$ would be a step towards smaller \K curvature. For example, with $3\a=1$ the \K curvature is  $-\mathcal{R}_K=2/3\a =2$. This is a smallest Poincar\'e disk associated with string theory. Smaller $\a$ are possible in $\N=1$ supergravity but the ones like $3\a=1, 2,3,4,5,6,7$ originate from string theory and maximal $\N=8$ supergravity.

It would be also interesting if the models with smaller values of $\m$ can be investigated, so that the system gets closer to the attractor regime and maybe works for smaller \K curvature.

\section{Discussion}

 All presently existing inflation-related data can be described by simple single-field inflationary models with one parameter  \cite{Kallosh:2021mnu,Kallosh:2022feu}.  Moreover, 
 the simple $\a$-attractor models  with $3\a=7,6,5,4,3,2,1$ form the set of discrete  B-mode targets, see   Fig. 2 in \cite{LiteBIRD:2022cnt}, where these 7 Poincar\'e disks are shown as some of the main targets for LiteBIRD. We  present  the predictions of the single stage inflationary $\a$-attractors in Fig. \ref{SnowmassPlus}.

Exponential $\a$-attractors make the nearly universal predictions $n_{s} =  1-{2/ N_{e}}$  and $r =12\alpha/N_{e}^{2}$. Using these  expressions one can derive a useful expression for the curvature of the \K geometry $\mathcal{R}_K= - {2\over 3\a}$:
\be
\mathcal{R}_K =  -{2(1-n_s)^2\over r}\, .
\label{simpleRK}
\ee
Thus by measuring $n_{s}$ and $r$ one can find the curvature  $\mathcal{R}_K$.

For polynomial attractors \cite{Kallosh:2022feu} this simple relation is no longer valid. Nevertheless, in this paper we argued that one can obtain  interesting information about the  curvature of the \K geometry $\mathcal{R}_K$ by investigation of  formation of PBHs and small-scale GWs. 

For single field exponential $\a$-attractors with an   inflection point, this possibility was studied in \cite{Dalianis:2018frf,Iacconi:2021ltm,Dalianis:2021dbs}. The current conclusion is that the PBHs produced in such models are light, $M_{\rm PBH} <10^8$ g,  due to the fact  that   the effective number of e-folds in such models is smaller than the one in models without the inflection point, and   $n_s$ cannot be smaller than the lowest value established by Planck \cite{Iacconi:2021ltm}. The peak of the GW signal is constrained to be at very high frequencies.
 
%Exponential $\a$-attractors make the nearly universal prediction $n_{s} =  1-{2\over N_{e}}$.  
In this paper we studied the  models  developed in \cite{Linde:2018hmx} and replaced 
there the exponential $\a$-attractor  part of these modes by a polynomial quadratic $\a$-attractor, which predicts a greater value of $n_{s} =1-{3\over 2N_{e}}$ \cite{Kallosh:2022feu}.  If there is a second stage of inflation driven by the axion field, it makes $N_e$ in  the attractor equation for $n_{s} $ smaller. The value of $n_{s}$ decreases, but if the second stage of inflation is not too long, $n_s$ decreases from its greater value $n_{s} =1-{3\over 2N_{e}}$  but still remains within the   Planck bounds on $n_s$. This tends to  remove the upper bound on the mass of PBHs of the type $M_{\rm PBH} <10^8$ g which was found in \cite{Iacconi:2021ltm}.

One of the most interesting and phenomenologically successful models of the PBHs production was proposed in \cite{Braglia:2020eai}, but its fundamental interpretation was not clear. We implemented this model in the context of polynomial $\a$-attractors, and developed its supergravity version. This embedding suggests a microscopic  origin of the parameters used in this model. In particular, the parameter $b_1$ in the axion kinetic term $e^{2 b_1 \phi} (\partial \chi)^2$  of \cite{Braglia:2020eai} depends on the curvature of the \K geometry, $b_1^2= -\mathcal{R}_K= {2\over 3 \a}$. 

The predictions of this model following from its polynomial attractor properties are:  $n_s$ and $r$ are $\a$-independent, whereas $r$ depends on the mass parameter $\mu$ defining  the  approach of the inflationary potential to the plateau. We  confirmed that the CMB predictions of this model are consistent with the attractor formula \rf{nsr1} for the polynomial $\a$-attractor model. We also explained a significant dependence of the PBH masses and GW frequencies on hyperbolic geometry curvature:   the effective tachyonic mass of isocurvature perturbations depends linearly on $\mathcal{R}_K$ at the point of transition from the dilaton to axion stage of inflation, as shown in eq. \rf{mass}.

%More experimental information on $n_s, r $ and PBHs and GWs will be needed to find out  which of the $\a$-attractor models can describe all  data: single field exponential or polynomial attractors with or without features, or two-stage inflationary models with features.

This suggests that by finding the spectrum of masses of PBHs  and the frequencies of the GWs, in combination with measuring $n_s, r$ one may find information about the  curvature of  the \K geometry $\mathcal{R}_K$  in this class of models. 
 
\section*{Acknowledgement}
We are grateful to  Y. Yamada  for useful comments on this work, and to A. Achucarro,   D.-G. Wang,   Y. Welling and Y. Yamada  for the collaboration on the earlier projects  which led to this paper. We are grateful to M. Sasaki and A. Starobinsky for the stimulating talks on inflation and PBHs production at the February 2022  Kyoto conference `Gravity - The Next Generation'.
  This work is  supported by SITP and by the US National Science Foundation Grant  PHY-2014215.

\appendix

\section{From half-plane to Poincar\'e disk coordinates}
To describe two-stage inflationary models based on hyperbolic geometry it is important to make a choice of geometric variables which are most convenient for understanding the cosmological evolution. In this paper we used  the half-plane coordinates  $T+\bar T>0$  with the complex field $T$ defining the dilaton-axion pair $(\vp, a)$
\be
T(x) = e^{- \sqrt{2\over 3\alpha} \vp (x) } + i \sqrt {2\over 3\alpha} \, a(x)  \ .
\label{T2}\ee
Alternatively, one could use the Poincar\'e disk coordinates $Z= r e^{i\theta} = {T-1\over T+1}$.

The map between the polar coordinates of the disk $(r, \theta)$  and planar coordinates of the hyperbolic field space $(\vp, a)$ is presented  in  \cite{Iacconi:2021ltm} with $\vp=u, a=v$. 
 \be
\vp(r, \theta) =\sqrt{3\alpha \over 2} \ln \Big[ {1+r^2+ 2 r \cos \theta\over 1-r^2}\Big ] \, , 
 \qquad a (r, \theta)= 2 \sqrt {3\alpha \over 2}  \Big[ {r \sin \theta\over 1+r^2+ 2 r \cos \theta}\Big ] 
\label{v} \ee
The relevant kinetic term in $(r, \theta)$ variables is
\be
-{3\a\over (1-r^2)^2} [(\partial r)^2 + r^2 (\partial \theta)^2] = -{1\over 2} (\partial \sigma)^2 - {3\a\over 4} \sinh^2 \Big (\sqrt{2\over 3\a} \sigma \Big) (\partial \theta)^2
\ee

The dilaton-axion potential which we use is given in  eq. \rf{Br}. We have shown it   in Fig. \ref{PotT2} together with the inflationary trajectory. %One can see that at the first stage the axion is not visibly moving, and when the dilaton stage ends, axion  starts.
To find the expression for this potential in  polar coordinates, one should put expressions    for  $\vp(r, \theta)$ and $a (r, \theta)$  in \rf{v}   into  the potential    \rf{Br}, and then replace the geometric variable $r$  by the canonically normalized field $\sigma$ such that  $r = \tanh  \sqrt{2\over 3\a} \sigma$,  $0\leq \sigma <\infty$.

\begin{figure}[H]
\centering
\includegraphics[scale=0.12]{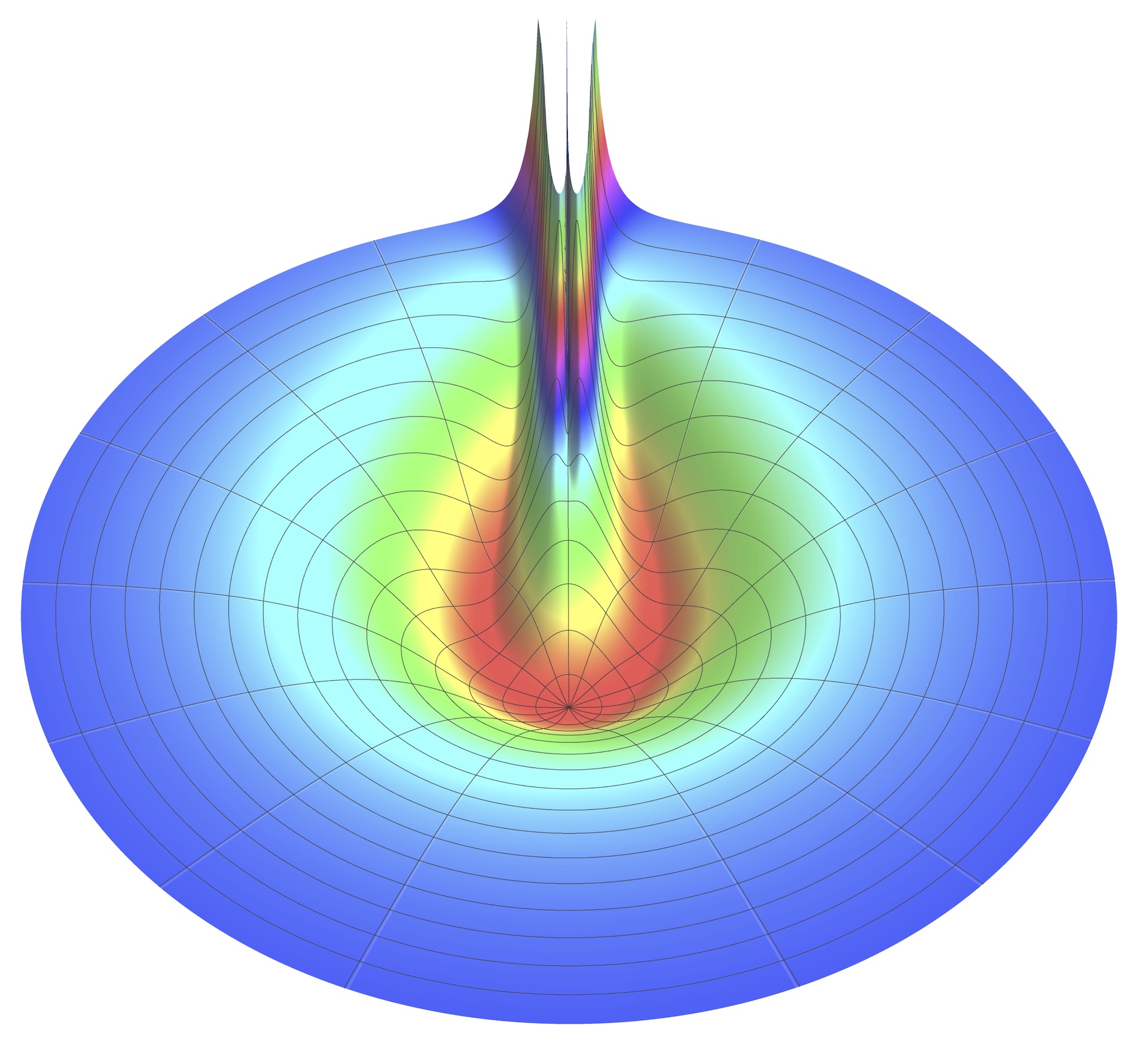}
\caption{\footnotesize The  potential  $V(\sigma, \theta)$. For the purpose of visualization we use $\a=1$ case in this plot. For much smaller values of $\alpha$ the ridges  become very narrow. The red area at the lower part  of the potentials corresponds to $\vp \approx 0$.  }
 \label{squids}
\end{figure}

%\be
%V_0 {\vp^2(r, \theta)\over \vp^2(r, \theta) + \m^2} +{1\over 2} m_a^2 a^2(r, \theta) \, , \qquad r=  \tanh  \sqrt{2\over 3\a} \sigma  \, , \qquad \sigma  \geq 0 \ . 
% \label{potPol}\ee
 We present the plot of the potential $V(\sigma, \theta)$ for  $\alpha = O(1)$ in  Fig. \ref{squids}.
 With some effort, one can figure out the behavior of the inflationary trajectory in $(\sigma, \theta)$ coordinates  in Fig.  \ref{squids}. In particular, the straight chaotic inflation valley at $\vp \approx 0$ shown as a red area in Fig. \ref{PotT2}  corresponds to the strongly curved red  valleys in Fig. \ref{squids}. This bending makes the description of the last stage of inflation slightly more complicated. For  smaller values of $\alpha$, such as $\alpha \sim 10^{-{2}}$ used in  \cite{Braglia:2020eai}, it is hard to draw and interpret figures like  Fig. \ref{squids}, because the ridges and the red valleys shown in these figures become extremely thin.

The potential  is singular at $r\to1$, $\theta \to \pi$  \cite{Iacconi:2021ltm}. The reason is that the axion field \rf{v} becomes infinitely large  in this limit, and the chaotic inflation potential  $
{1\over 2} m_a ^2  a^2
$ diverges in this limit $a\to \infty$. This is the standard feature of  monomial chaotic inflation potentials. This singularity disappears if one replaces  $
{1\over 2} m_a ^2  a^2$ by a periodic axion potential as in \cite{Achucarro:2017ing,Linde:2018hmx}.

%However, cosmology is much simpler in Fig.  \ref{PotT}  in dilaton-axion coordinates. 

Thus,  it is possible to change coordinates in the moduli space by a Cayley transform, $T= {1+Z\over 1-Z}, \, Z= {T-1\over T+1}$, as shown in Figs. \ref{Escher2022}, \ref{PotT2}, \ref{squids}.   It is amazing that we can map the behavior of the axion potential in the infinite range $-\infty < a < +\infty$ to the vicinity of the disc boundary.  However,  the cosmological two-stage inflation models, which we study in this paper, look much  simpler in the dilaton-axion form \rf{T}, \rf{full} and in Fig. \ref{PotT2}.

\bibliographystyle{JHEP}
\bibliography{lindekalloshrefs}
\end{document}